\documentclass[epj]{webofc}
\usepackage[utf8]{inputenc}
\usepackage[varg]{txfonts}   
\usepackage{booktabs}
\usepackage{xcolor}
\definecolor{darkred}{rgb}{0.4,0.0,0.0}
\definecolor{darkgreen}{rgb}{0.0,0.4,0.0}
\definecolor{darkblue}{rgb}{0.0,0.0,0.4}
\usepackage[bookmarks,linktocpage,colorlinks,
    linkcolor = darkred,
    urlcolor  = darkblue,
    citecolor = darkgreen]{hyperref}
\wocname{EPJ Web of Conferences}
\woctitle{Lattice2017}
\usepackage{graphicx}
\usepackage{amssymb}
\usepackage{epstopdf}

\newcommand{\be}{\begin{equation}}
\newcommand{\ee}{\end{equation}}
\newcommand{\bea}{\begin{eqnarray}}
\newcommand{\eea}{\end{eqnarray}}
\begin{document}
%
\selectlanguage{english}
\title{
HVP contributions to the muon ($g - 2$) including QED corrections with twisted-mass fermions\thanks{Presented at the XXXV Int'l Symposium on Lattice Field Theory, Granada (Spain), June 18-24, 2017.}
}
\author{
\firstname{Davide} \lastname{Giusti}\inst{1,2} \and
\firstname{Vittorio} \lastname{Lubicz}\inst{2} \and
\firstname{Guido} \lastname{Martinelli}\inst{3} \and
\firstname{Francesco} \lastname{Sanfilippo}\inst{2}  \and
\firstname{Silvano} \lastname{Simula}\inst{2}
}
\institute{
Dip.~di Matematica e Fisica, Universit\`a di Roma Tre, Via della Vasca Navale 84, I-00146 Rome, Italy
\and
Istituto Nazionale di Fisica Nucleare, Sezione di Roma Tre, Via della Vasca Navale 84, I-00146 Rome, Italy
\and
Dip.~di Fisica, Universit\`a di Roma ``La Sapienza'' and INFN Sezione di Roma, Piazzale Aldo Moro 5, 00185 Roma, Italy
}
\abstract{
We present a lattice calculation of the Hadronic Vacuum Polarization (HVP) contribution of the strange and charm quarks to the anomalous magnetic moment of the muon including leading-order electromagnetic (e.m.) corrections. 
We employ the gauge configurations generated by the European Twisted Mass Collaboration (ETMC) with $N_f = 2+1+1$ dynamical quarks at three values of the lattice spacing ($a \simeq 0.062, 0.082, 0.089$ fm) with pion masses in the range $M_\pi \simeq 210 - 450$ MeV. 
The strange and charm quark masses are tuned at their physical values. 
Neglecting disconnected diagrams and after the extrapolations to the physical pion mass and to the continuum limit we obtain: $a_\mu^s(\alpha_{em}^2) = (53.1 \pm 2.5) \cdot 10^{-10}$, $a_\mu^s(\alpha_{em}^3) = (-0.018 \pm 0.011) \cdot 10^{-10}$ and $a_\mu^c(\alpha_{em}^2) = (14.75 \pm 0.56) \cdot 10^{-10}$, $a_\mu^c(\alpha_{em}^3) = (-0.030 \pm 0.013) \cdot 10^{-10}$ for the strange and charm contributions, respectively. 
}
\maketitle
\section{Introduction}\label{intro}

The  anomalous magnetic moment of the muon $a_\mu \equiv (g -2 ) / 2$ is known experimentally with an accuracy of the order of 0.54 ppm, while the current precision of the Standard Model (SM) prediction is at the level of 0.4 ppm~\cite{PDG}.
The tension of the experimental value with the SM prediction, $a_\mu^{exp} - a_\mu^{SM} = (28.8 \pm 8.0) \cdot 10^{-10}$ \cite{PDG}, corresponds to $\simeq 3.5$ standard deviations and might be an exciting indication of new physics.
The forthcoming $g - 2$ experiments at Fermilab (E989)~\cite{Logashenko:2015xab} and J-PARC (E34)~\cite{Otani:2015lra} aim at reducing the experimental uncertainty by a factor of four, down to 0.14 ppm.
Such a precision makes the comparison of the experimental value of $a_\mu$ with theoretical predictions one of the most important tests of the Standard Model in the quest for new physics effects.
 
It is clear that the experimental precision must be matched by a comparable theoretical accuracy.
With a reduced experimental error, the uncertainty of the hadronic corrections will soon become the main limitation of this test of the SM. 
For this reason an intense research program is under way to improve the evaluation of the leading-order hadronic contribution to $a_\mu$ due to the HVP correction to the one-loop diagram, $a_\mu^{had}(\alpha_{em}^2)$, as well as to the next-to-leading-order hadronic corrections, which include $O(\alpha_{em}^3)$ contributions (see Ref.~\cite{Jegerlehner:2009ry}).  

The theoretical predictions for the hadronic contributions are traditionally obtained using dispersion relations for relating the HVP term to the experimental cross section data for $e^+ e^-$ annihilation into hadrons \cite{Davier:2010nc,Hagiwara:2011af}. 
An alternative approach, proposed in Refs.~\cite{Lautrup:1971jf,deRafael:1993za,Blum:2002ii}, is to compute $a_\mu^{had}(\alpha_{em}^2)$ in Euclidean lattice QCD from the correlation function of two e.m.~currents. 
In this respect an impressive progress in the lattice determinations of $a_\mu^{had}(\alpha_{em}^2)$ has been achieved in the last few years \cite{Boyle:2011hu,DellaMorte:2011aa,Burger:2013jya,Chakraborty:2014mwa,Chakraborty:2015cso,Bali:2015msa,Chakraborty:2015ugp,Blum:2015you,Blum:2016xpd,Chakraborty:2016mwy,DellaMorte:2017dyu}. 

With the increasing precision of the lattice calculations, it becomes necessary to include e.m.~and strong isospin breaking (IB) corrections (contributing at order $O(\alpha_{em}^3)$ and $O(\alpha_{em}^2 (m_d - m_u))$, respectively) to the  HVP.  
In this contribution we present the results of a lattice calculation of the e.m.~corrections to the HVP contribution due to strange and charm quark intermediate states, obtained in Ref.~\cite{Giusti:2017jof} using the expansion method of Refs.~\cite{deDivitiis:2011eh,deDivitiis:2013xla}. 
Given the large statistical fluctuations, we will show only very preliminary results for the e.m.~and IB corrections to the HVP contribution due to up and down quarks.  
For the same reason we do not have yet results for the disconnected contributions.

\section{Master formula}\label{sec:master}

The hadronic contribution $a_\mu^{had}$ to the muon anomalous magnetic moment at order $\alpha_{em}^2$ can be related to the Euclidean space-time HVP function $\Pi(Q^2)$ by~\cite{Lautrup:1971jf,deRafael:1993za,Blum:2002ii}
 \be
      a_\mu^{had} = 4 \alpha_{em}^2 \int_0^\infty dQ^2 f(Q^2) \left[ \Pi(Q^2) -  \Pi(0) \right] ~ ,
      \label{eq:amu}
 \ee
where $Q$ is the Euclidean four-momentum and $f(Q^2)$ is a well-know kinematical kernel, depending also on the muon mass $m_\mu$.
The HVP function $[\Pi(Q^2) - \Pi(0)]$ can be determined from the vector current-current Euclidean correlator $V(t)$ defined as
 \be
     V(t) \equiv \frac{1}{3} \sum_{i=1,2,3} \int d\vec{x} ~ \langle J_i(\vec{x}, t) J_i(0) \rangle ~ ,
     \label{eq:VV}
 \ee
 where $J_\mu = \sum_{f = u, d, s, c, ...} q_f ~ \overline{\psi}_f(x) \gamma_\mu \psi_f(x) $ is the e.m.~current with $q_f$ being the electric charge of the quark with flavor $f$ in units of $e$.
One gets \cite{Bernecker:2011gh}
 \be
      a_\mu^{had} = 4 \alpha_{em}^2 \int_0^\infty dt ~ \tilde{f}(t) V(t) ~ ,
      \label{eq:amu_t}
 \ee
 where $\tilde{f}(t)$ is given by
  \be
      \tilde{f}(t) \equiv 2 \int_0^\infty dQ^2 ~ f(Q^2) \left[ \frac{\mbox{cos}(Qt) -1}{Q^2} + \frac{1}{2} t^2 \right] 
      \label{eq:ftilde}
  \ee
and can be easily calculated at any value of $t$.
In what follows we will limit ourselves to the connected contributions to $a_\mu^{had}$.
In this case each quark flavor $f$ contributes separately.

The vector correlator $V(t)$ can be calculated on a lattice with volume $L^3$ and temporal extension $T$ at discretized values of $\overline{t} \equiv t / a$ from $0$ to $\overline{T}/2$ with $\overline{T} = T / a$.
In what follows all the overlined quantities are in lattice units.
A natural procedure is to split Eq.~(\ref{eq:amu_t}) into two contributions corresponding to $0 \leq \overline{t} \leq \overline{T}_{data}$ and $\overline{t} > \overline{T}_{data}$, respectively.
In the first contribution the vector correlator is directly given by the lattice data, while for the second contribution an analytic representation is required (see Refs.~\cite{Chakraborty:2015ugp,Blum:2015you,Chakraborty:2016mwy,DellaMorte:2017dyu}).
If $\overline{T}_{data}$ is large enough that the ground-state contribution is dominant for $\overline{t} > \overline{T}_{data}$, one can write
\be
     a_\mu^{had} = 4 \alpha_{em}^2 \left\{ ~ \sum_{\overline{t} = 0}^{\overline{T}_{data}} 
                             \overline{f}(\overline{t}) \overline{V}(\overline{t}) + 
                             \sum_{\overline{t} = \overline{T}_{data} + 1}^\infty \overline{f}(\overline{t}) 
                            \frac{\overline{Z}_V}{2 \overline{M}_V} e^{- \overline{M}_V \overline{t}}  ~ \right\} ~ ,
     \label{eq:amuL}
 \ee
where $\overline{Z}_V \equiv (1/3) \sum_{i=1,2,3} | \langle 0| J_i(0) | V \rangle |^2$ is the (squared) matrix element of the current operator between the vector ground-state and the vacuum. 
For each gauge ensemble the masses $\overline{M}_V$ and the matrix elements $\overline{Z}_V$ are extracted from a single exponential fit of the vector correlator $V(t)$ in the range $\overline{t}_{min} \leq \overline{t} \leq \overline{t}_{max}$, given explicitly in Ref.~\cite{Giusti:2017jof}.

\section{Simulation details}
\label{sec:simulations}

The ETMC gauge ensembles used in this contribution are the same adopted in Ref.~\cite{Carrasco:2014cwa} to determine the up, down, strange and charm quark masses. 
We employed the Iwasaki action for gluons  and the Wilson Twisted Mass Action for sea quarks. 
In order to avoid the mixing of strange and charm quarks in the valence sector we adopted a non-unitary set up in which the valence strange and charm quarks are regularized as Osterwalder-Seiler fermions, while the valence up and down quarks have the same action of the sea.
Working at maximal twist such a setup guarantees an automatic ${\cal{O}}(a)$-improvement.

We considered three values of the inverse bare lattice coupling $\beta$ and different lattice volumes.
At each lattice spacing, different values of the light sea quark masses have been considered. 
The light valence and sea quark masses are always taken to be degenerate. 
The bare masses of both the strange ($a\mu_s$) and the charm ($a\mu_c$) valence quarks are obtained, at each $\beta$, using the physical strange and charm masses and the mass renormalization constant (RC) determined in Ref.~\cite{Carrasco:2014cwa}.
The values of the lattice spacing are: $a = 0.0885(36)$, $0.0815(30)$, $0.0619(18)$ fm at $\beta = 1.90$, $1.95$ and $2.10$, respectively.

We made use of the bootstrap samplings elaborated for the input parameters of the quark mass analysis of Ref.~\cite{Carrasco:2014cwa}.
There, eight branches of the analysis were adopted differing in: ~ i) the continuum extrapolation adopting for the scale parameter either the Sommer parameter $r_0$ or the mass of a fictitious PS meson made up of strange(charm)-like quarks; ~ ii) the chiral extrapolation performed with fitting functions chosen to be either a polynomial expansion or a Chiral Perturbation Theory Ansatz in the light-quark mass; and ~ iii) the choice between the methods M1 and M2, which differ by $O(a^2)$ effects, used to determine in the RI'-MOM scheme the mass RC $Z_m = 1 / Z_P$.

In our numerical simulations the evaluation of the vector correlator has been carried out using the following local current:
 \be
    J_\mu(x) = Z_A ~ q_f ~ \bar{\psi}_{f^\prime}(x) \gamma_\mu \psi_f(x) ~ ,
    \label{eq:localV}
 \ee
where $\psi_{f^\prime}$ and $\psi_f$ represent two quarks with the same mass and charge, but regularized with opposite values of the Wilson $r$-parameter, i.e.~$r_{f^\prime} = - r_f$.
Being at maximal twist the current (\ref{eq:localV}) renormalizes multiplicatively with the RC $Z_A$ of the axial current.
The choice (\ref{eq:localV}) is characterized by the absence of disconnected insertions (see Refs.~\cite{Chakraborty:2015ugp,Blum:2015you,Giusti:2017jof}).

The statistical accuracy of the meson correlators is based on the use of the so-called ``one-end" stochastic method \cite{McNeile:2006bz}, which includes spatial stochastic sources at a single time slice chosen randomly.
Four stochastic sources (diagonal in the spin variable and dense in the color one) were adopted per each gauge configuration.

\section{Lowest order}
\label{sec:s&c}

For the evaluation of the strange and charm contributions to $a_\mu^{had}$ we have adopted four choices of $\overline{T}_{data}$, namely: $\overline{T}_{data} = (\overline{t}_{min}+2)$, $(\overline{t}_{min} + \overline{t}_{max}) / 2$, $(\overline{t}_{max} - 2)$ and $(\overline{T} / 2 - 4)$.
In Ref.~\cite{Giusti:2017jof} it is shown that $a_\mu^{had}$ is almost independent of the specific choice of the value of $\overline{T}_{data}$.

\begin{figure}[htb!]
\centering{\scalebox{0.725}{\includegraphics{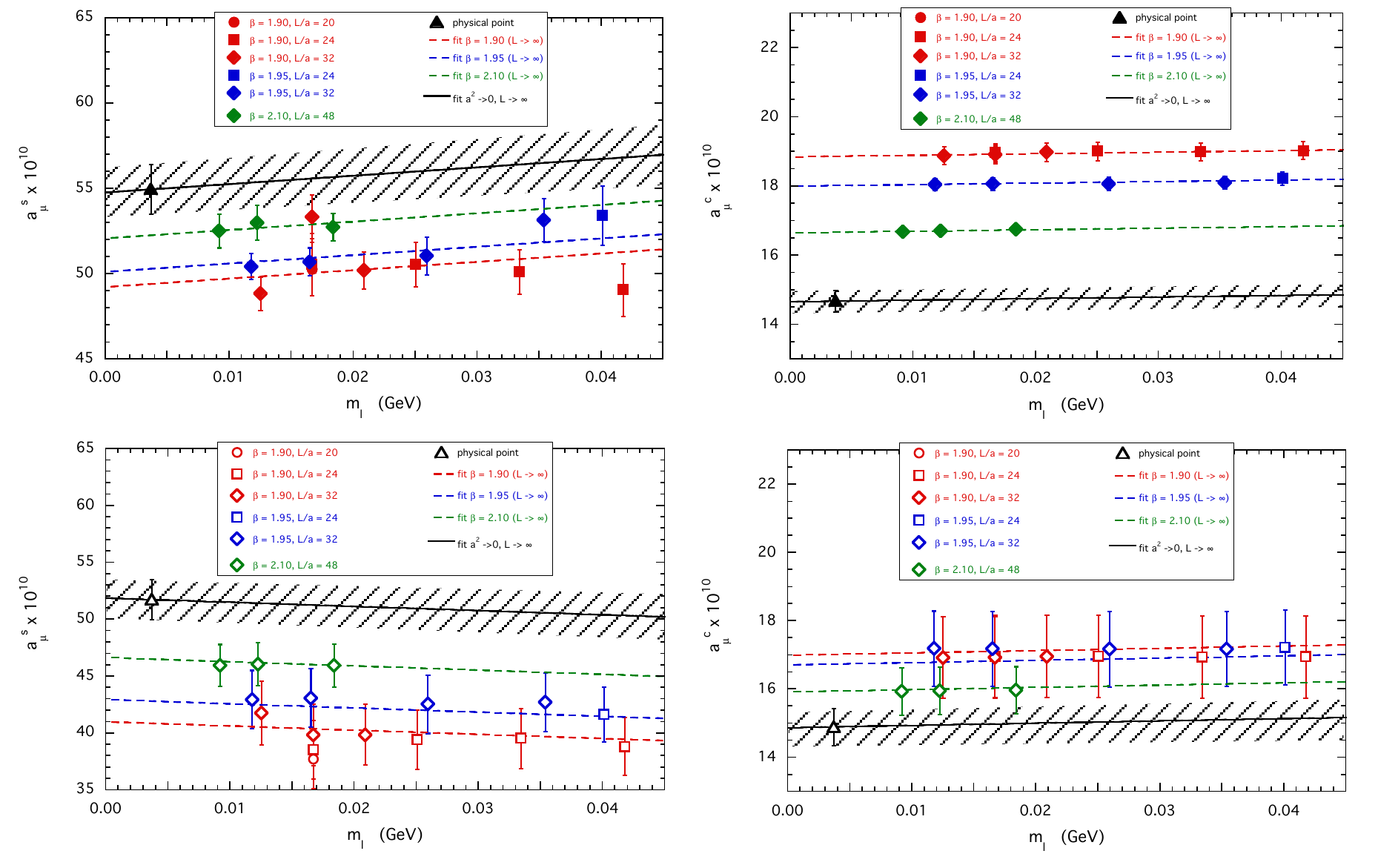}}}
\vspace{-0.5cm}
\caption{\it \small Results for the strange (left panels) and charm (right panels) contributions to $a_\mu^{had}$ in units of $10^{-10}$. Upper (lower) panels correspond to the data obtained with (without) the ELM procedure (\ref{eq:HLtrick}). The dashed lines correspond to the linear fit (\ref{eq:fit_amus}) in the infinite volume limit. The solid lines correspond to the continuum and infinite volume limits, while the shaded areas identify the uncertainty at $1 \sigma$ level. The triangles are the results of the extrapolation at the physical pion mass and in the continuum and infinite volume limits.}
\label{fig:amuq_fit}
\end{figure}

The results obtained for the strange and charm contributions to $a_\mu^{had}$ are shown by the empty markers in the lower panels of Fig.~\ref{fig:amuq_fit}.
We observe a mild dependence on the light-quark mass, being driven only by sea quarks, and also small residual finite size effects (FSEs) are visible only in the case of the strange contribution.
The errors of the data turn out to be dominated by the uncertainties of the scale setting, which are similar for all the gauge ensembles used in this contribution.

In Ref.~\cite{Burger:2013jya} a modification of $a_\mu^{had}$ at pion masses above the physical point has been proposed in order to weaken its pion mass dependence and improve the reliability of the chiral extrapolation. 
Though the procedure of Ref.~\cite{Burger:2013jya} has been conceived mainly for the light contribution to $a_\mu^{had}$, we have explored its usefulness also in the case of the strange and charm contributions.
The proposal consists in multiplying the Euclidean 4-momentum transfer $Q^2$ by a factor equal to $(M_V / M_V^{phys})^2$ in order to modify the $Q^2$-dependence of the HVP function $\Pi_R(Q^2)$ without modifying its value at the physical point.
One obtains the same effect by redefining the lepton mass as
 \be
    \overline{m}_\mu^{ELM} = m_\mu ~ \overline{M}_V / M_V^{phys} ~ .
    \label{eq:HLtrick}
 \ee
The expected advantage of the use of the effective lepton mass (\ref{eq:HLtrick}) comes from the fact that the kernel function $\overline{f}(\overline{t})$, and therefore $a_\mu^{had}$, depends only on the lepton mass in lattice units.
Thanks to Eq.~(\ref{eq:HLtrick}), which will be referred to as the Effective Lepton Mass (ELM) procedure, the knowledge of the value of the lattice spacing is not required and therefore the resulting $a_\mu^{had}$ is not affected by the uncertainties of the scale setting.
The drawback of the ELM procedure is instead represented by its potential sensitivity to the statistical fluctuations of the vector meson mass extracted from the lattice data.

The results obtained adopting the ELM procedure (\ref{eq:HLtrick}) in the case of the strange and charm contributions to $a_\mu^{had}$ are shown by the filled markers in Fig.~\ref{fig:amuq_fit}, where the physical values for the $\bar{s} s$ and $\bar{c} c$ vector masses have been taken from PDG \cite{PDG} (namely, $M_V^{(phys)} = 1.0195$ and $3.0969$ GeV, respectively).
It can be seen that the ELM procedure reduces remarkably the overall uncertainty of the data.
Moreover, it further weakens the pion mass dependence (in any case driven only by the sea quarks) and modifies the discretization effects, leading to a better scaling behavior of the data in the case of the charm contribution.
Since the pion mass dependence is in any case quite mild, the ELM procedure can be viewed as an alternative way to perform the continuum extrapolation and to avoid the scale setting uncertainties.

We have performed a combined fit for the extrapolation to the physical pion mass, the continuum and infinite volume limits using the following Ansatz
 \be
     a_\mu^{s,c} = A_0^{s,c} \left[ 1 + A_1^{s,c} \xi + D^{s,c} a^2 + F^{s,c} \xi e^{-M_\pi L} /(M_\pi L) \right] ~ ,
     \label{eq:fit_amus}
  \ee
where $\xi \equiv M_\pi^2 / (4 \pi f_0)^2$ and the exponential term is a phenomenological representation of possible FSEs.
The results of the linear fit (\ref{eq:fit_amus}) are shown in Fig.~\ref{fig:amuq_fit} by the solid lines.

Averaging over the results corresponding to different fitting functions of the data either with or without the ELM procedure we get at the physical point
 \bea
     a_\mu^{s, phys} & = & (53.1 \pm 1.6_{stat+fit} \pm 1.5_{input} \pm 1.3_{disc} \pm 0.2_{FSE}
                                         \pm 0.1_{chir}) \cdot 10^{-10} ~ , \nonumber \\
                               & = & (53.1 \pm 1.6_{stat+fit} \pm 2.0_{syst}) \cdot 10^{-10} 
                                  = (53.1 \pm 2.5) \cdot 10^{-10} ~ ,
     \label{eq:amus_phys}
 \eea
where
\begin{itemize}
\item $()_{stat+fit}$ indicates indicates the uncertainty induced by both the statistical errors and the fitting procedure itself;
\item $()_{input}$ is the error coming from the uncertainties of the input parameters of the eight branches of the quark mass analysis of Ref.~\cite{Carrasco:2014cwa};
\item $()_{disc}$ is the uncertainty due to both discretization effects and scale setting, estimated by comparing the results obtained with and without the ELM procedure (\ref{eq:HLtrick});
\item $()_{FSE}$ is the error coming from including ($F^s \neq 0$) or excluding ($F^s = 0$) the FSE correction. When FSEs are not included, all the gauge ensembles with $L / a = 20$ and $24$ are also not included;
\item $()_{chir}$ is the error coming from including ($A_1^s \neq 0$) or excluding ($A_1^s = 0$) the linear term in the light-quark mass.
\end{itemize}

Our result (\ref{eq:amus_phys}) compares well with the $N_f = 2+1+1$  result $a_\mu^{s, phys} = (53.41 \pm 0.59) \cdot 10^{-10}$ from HPQCD~\cite{Chakraborty:2014mwa}, the $N_f = 2+1$ finding $a_\mu^{s, phys} = (53.1 \pm 0.9_{-0.3}^{+0.1}) \cdot 10^{-10}$ obtained by RBC/UKQCD~\cite{Blum:2016xpd}, and with the recent $N_f = 2$ result $a_\mu^{s, phys} = (51.1 \pm 1.7 \pm 0.4) \cdot 10^{-10}$ of Ref.~\cite{DellaMorte:2017dyu}.

In the case of the charm contribution we obtain
 \bea
     a_\mu^{c, phys} & = & (14.75 \pm 0.42_{stat+fit} \pm 0.36_{input} \pm 0.10_{disc} \pm 0.03_{FSE} \pm 0.01_{chir}) \cdot 10^{-10} ~ , \nonumber \\
                                & = & (14.75 \pm 0.42_{stat+fit} \pm 0.37_{syst}) \cdot 10^{-10} 
                                    =  (14.75 \pm 0.56) \cdot 10^{-10} ~ ,
     \label{eq:amuc_phys}
 \eea
where the errors are estimated as in the case of the strange quark contribution.
Our finding (\ref{eq:amuc_phys}) agrees with the $N_f = 2+1+1$ result $a_\mu^{c, phys} = (14.42 \pm 0.39) \cdot 10^{-10}$ from HPQCD~\cite{Chakraborty:2014mwa} and with recent $N_f = 2$ one $a_\mu^{c, phys} = (14.3 \pm 0.2 \pm 0.1) \cdot 10^{-10}$ of Ref.~\cite{DellaMorte:2017dyu}.

\section{Electromagnetic corrections}
\label{sec:deltas&c}

Let's now turn to the e.m.~correction $\delta V(t)$ to the vector correlator at leading order in $\alpha_{em}$.
Using the expansion method of Ref.~\cite{deDivitiis:2013xla} for each quark flavor $f$ it can be written as
 \be
     \delta V(t) \equiv \delta V^{self}(t) + \delta V^{exch}(t) + \delta V^{tad}(t) + \delta V^{PS}(t) + \delta V^S(t) ~ 
     \label{eq:deltaV}
 \ee
where the various terms correspond to the evaluation of the self-energy, exchange, tadpole, pseudoscalar and scalar insertion diagrams (see Fig.~8 of Ref.~\cite{Giusti:2017jof}).
The removal of the photon zero-mode is done according to $QED_L$~\cite{Hayakawa:2008an}, i.e.~the photon field $A_\mu$ satisfies $A_\mu(k_0, \vec{k} = \vec{0}) \equiv 0$ for all $k_0$.
 
In addition one has to consider the QED contribution to the RC $Z_A$ of the vector current (\ref{eq:localV}):
 \be
     Z_A = Z_A^{(0)} \left[ 1 + Z_A^{(em)} ~ Z_A^{(fact)} \right] + {\cal{O}}(\alpha_{em}^2) ~ ,
     \label{eq:ZA}
 \ee
where $Z_A^{(0)}$ is the RC in absence of QED (determined in Ref.~\cite{Carrasco:2014cwa}), $Z_A^{(em)}$ is the one-loop perturbative estimate of the QED effect at order ${\cal{O}}(\alpha_s^0)$ and $Z_A^{(fact)}$ takes into account corrections of order ${\cal{O}}(\alpha_{em} \alpha_s^n)$ with $n \geq 1$, i.e.~corrections to the ``naive factorization approximation" in which $Z_A^{(fact)} = 1$.
In Ref.~\cite{Giusti:2017jof} the non-perturbative estimate $Z_A^{(fact)} = 0.9 \pm 0.1$ has been obtained through the use of the axial Ward-Takahashi identity (WTI) derived in the presence of QED effects.
Using the result $Z_A^{(em)} = - 15.7963 ~ \alpha_{em} ~ q_f^2 / (4 \pi)$ from Refs.~\cite{Martinelli:1982mw}, we have to add to Eq.~(\ref{eq:deltaV}) the following contribution
 \be
      \delta V^{Z_A}(t) \equiv -2.51406 ~ \alpha_{em} q_f^2 ~ Z_A^{(fact)} ~ V(t) ~ .
      \label{eq:deltaV_ZA}
 \ee

Thus, the e.m.~corrections $\delta a_\mu^{had}$ can be written as
\be
     \delta a_\mu^{had} = 4 \alpha_{em}^2 \left\{ ~ \sum_{\overline{t} = 0}^{\overline{T}_{data}} 
                                       \overline{f}(\overline{t}) ~ \delta \overline{V}(\overline{t}) + 
                                        \sum_{\overline{t} = \overline{T}_{data} + 1}^\infty 
                                        \overline{f}(\overline{t}) ~ \frac{\overline{Z}_V} {2 \overline{M}_V} 
                                        e^{- \overline{M}_V \overline{t}} \left[ \frac{\delta \overline{Z}_V}{\overline{Z}_V}
                                        - \frac{\delta \overline{M}_V}{\overline{M}_V} (1 + \overline{M}_V \overline{t}) 
                                        \right] ~ \right\} ~ ,
     \label{eq:deltamu}
 \ee
where $\delta \overline{M}_V$ and $\delta \overline{Z}_V$ can be determined, respectively, from the ``slope'' and the ``intercept'' of the ratio $\delta \overline{V}(\overline{t}) / \overline{V}(\overline{t})$ at large time distances $\overline{t}_{min} \leq \overline{t} \leq \overline{t}_{max}$ (see Refs.~\cite{deDivitiis:2011eh,deDivitiis:2013xla,Giusti:2017dmp}).

\begin{figure}[htb!]
\centering{\scalebox{0.725}{\includegraphics{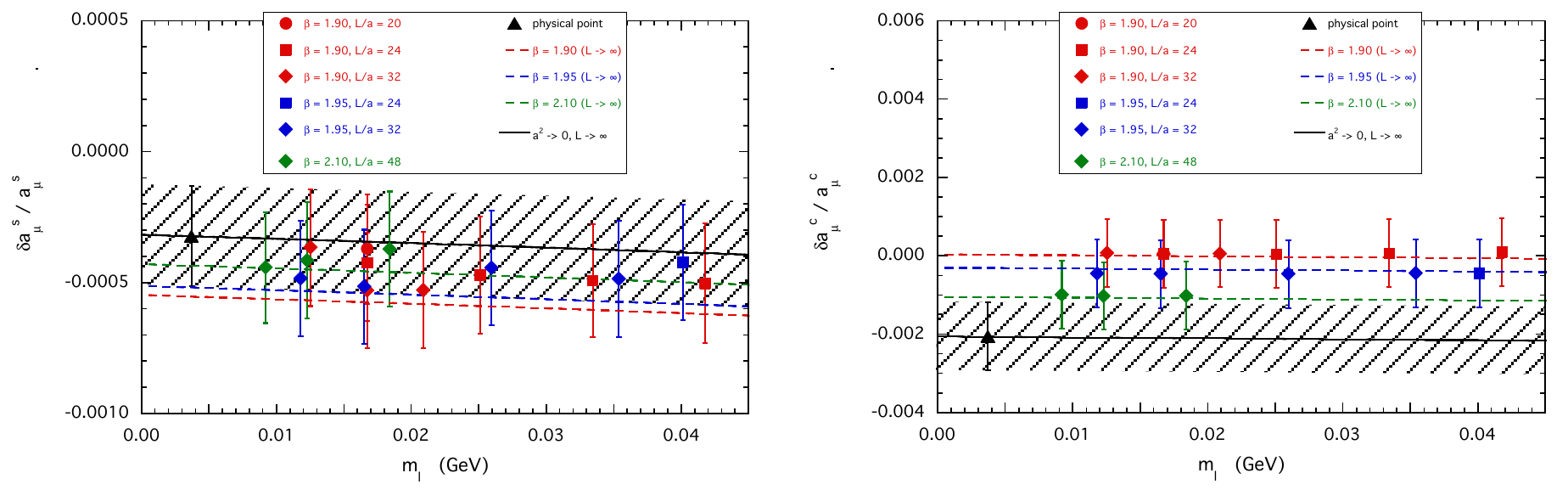}}}
\vspace{-0.5cm}
\caption{\it \small Results for the strange (left panel) and charm (right panel) contributions to $\delta a_\mu^{had} / a_\mu^{had}$. The dashed lines correspond to the linear fit (\ref{eq:fit_damuq}) in the infinite volume limit. The solid lines correspond to the continuum and infinite volume limits, while the shaded areas identify the uncertainty at $1 \sigma$ level. The triangles are the results of the extrapolation at the physical pion mass and in the continuum and infinite volume limits.}
\label{fig:ratioq_fit}
\end{figure}

As in the case of the lowest-order terms $a_\mu^{had}$, we adopt for the evaluation of $\delta a_\mu^{had}$ the same four choices of $\overline{T}_{data}$.
We find that $\delta a_\mu^{had}$ is largely independent of the choice of the value of $\overline{T}_{data}$ within the statistical uncertainties.
In the case of the e.m.~corrections the use of the ELM procedure (\ref{eq:HLtrick}) does not improve the precision of the lattice data.
Instead, this can be achieved by forming the ratio $\delta a_\mu^{had} / a_\mu^{had}$.
The results for the latter are shown in Fig.~\ref{fig:ratioq_fit}.

It can be seen that the dependence on the light-quark mass $m_\ell$ is quite mild, being driven only by sea quarks, and that the uncertainties of the data are dominated by the error on the RC $Z_A^{fact}$, which has been taken to be the same for all the gauge ensembles used in this contribution.

The FSEs are visible only in the case of the strange quark.
A theoretical calculation of FSEs for $\delta a_\mu^{had}$ is not yet available.
According to Ref.~\cite{Lubicz:2016xro} the universal FSEs are expected to vanish, since they depend on the global charge of the meson states appearing in the spectral decomposition of the correlator $\delta V(t)$.
Moreover, the structure-dependent (SD) FSEs are expected to start at order ${\cal{O}}(1/L^2)$.
According to Ref.~\cite{Davoudi:2014qua} one might argue that in the case of mesons with vanishing charge radius the SD FSEs may start at order ${\cal{O}}(1/L^3)$.
Therefore we adopt the following simple fitting function
 \be
     \delta a_\mu^{s,c} / a_\mu^{s,c} = \delta A_0^{s,c} + \delta A_1^{s,c} m_\ell + \delta D^{s,c} a^2 + 
         \delta F^{s,c} / L^n 
     \label{eq:fit_damuq}
 \ee
where the power $n$ can be put equal to $n = 2$ or $n = 3$.
In fitting our data we do not observe sensitivity to the above choices of the power $n$ within the statistical uncertainties.

At the physical pion mass and in the continuum and infinite volume limits we get
 \bea
      \label{eq:damus_phys}
     \frac{\delta a_\mu^{s, phys}}{a_\mu^{s, phys}} & = & -0.000332 ~ (46)_{stat+fit} ~ (6)_{input} ~ 
         (8)_{FSE} ~ (4)_{chir} ~ (2)_{disc} ~ (208)_{Z_A} ~ , \nonumber \\
                                                                               & = & -0.000332 ~ (46)_{stat+fit} ~ (208)_{syst} 
                                                                                  = -0.000332 ~ (213) ~ ,  \\[2mm] 
    \label{eq:damuc_phys}  
     \frac{\delta a_\mu^{c, phys}}{a_\mu^{c, phys}} & = & -0.00205 ~ (12)_{stat+fit} ~ (1)_{input} ~ 
         (1)_{FSE} ~ (1)_{chir} ~ (1)_{disc} ~ (85)_{Z_A} ~ , \nonumber \\
                                                                               & = & -0.00205 ~ (12)_{stat+fit} ~ (85)_{syst} 
                                                                                  = -0.00205 ~ (86) ~ ,
 \eea
where the error budget is similar to the one described for the lowest-order results, while $()_{Z_A}$ is the error generated by the uncertainty on the RC $Z_A$ (see Eq.~(\ref{eq:ZA})).
The latter one is by far the dominant source of uncertainty.
Using the lowest-order results (\ref{eq:amus_phys}-\ref{eq:amuc_phys}) we obtain 
 \be
     \label{eq:damusc_phys}
     \delta a_\mu^{s, phys} = - 0.018 ~ (11) \cdot 10^{-10} ~ , \qquad
     \delta a_\mu^{c, phys} = - 0.030 ~ (13) \cdot 10^{-10} ~ ,   
 \ee
showing that the e.m.~corrections $\delta a_\mu^s$ and $\delta a_\mu^c$ are negligible with respect to the uncertainties of the lowest-order terms.
We stress that the errors appearing in Eq.~(\ref{eq:damusc_phys}) are dominated by the uncertainty on the RC $Z_A$ of the local vector current, estimated through the axial WTI in the presence of QED effects.
A dedicated study aimed at the determination of the RCs of bilinear operators in the presence of QED employing non-perturbative renormalization schemes, like the RI-MOM one, is expected to improve the precision of the calculation of the e.m.~corrections and IB effects on $a_\mu^{had}$.

Our findings demonstrate that the expansion method of Ref.~\cite{deDivitiis:2013xla}, already applied successfully to the calculation of the e.m.~corrections to meson masses~\cite{deDivitiis:2013xla,Giusti:2017dmp} and to the leptonic decays of pions and kaons \cite{Lubicz:2016mpj,Tantalo:2016vxk}, works as well also in the case of the HVP contribution to the muon $(g -2)$.

\begin{figure}[htb!]
\centering{\scalebox{0.725}{\includegraphics{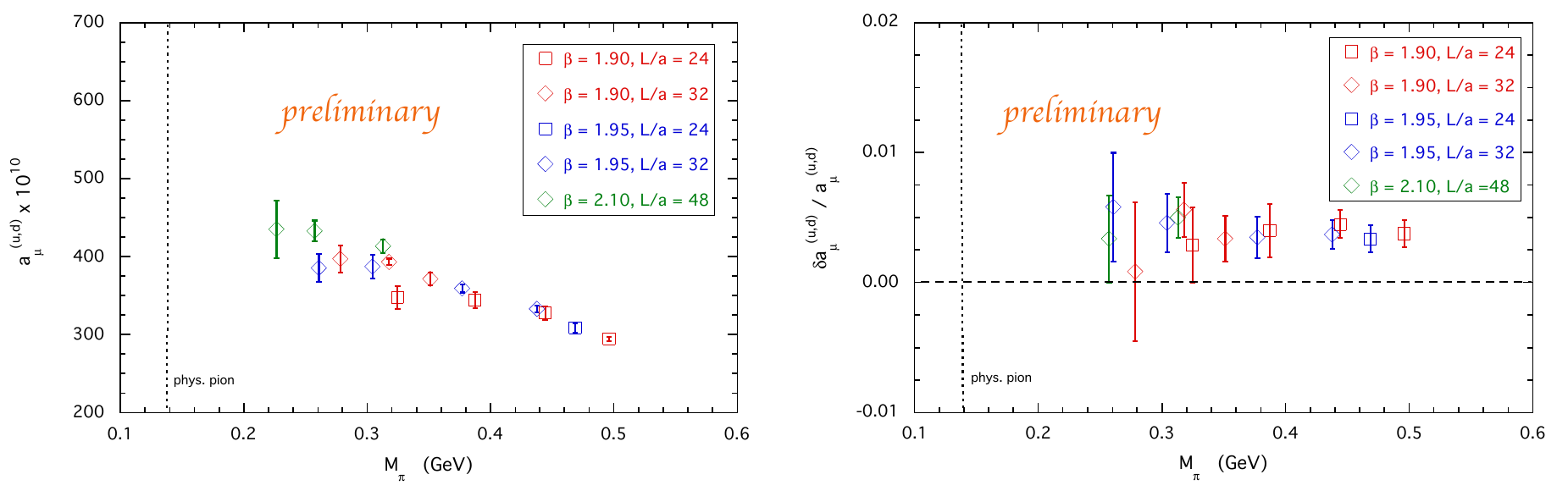}}}
\vspace{-0.5cm}
\caption{\it \small Preliminary results for the $u$ and $d$-quark contributions $a_\mu^{(u,d)}$ (left panel) and $\delta a_\mu^{(u,d)} / a_\mu^{(u,d)}$ (right panel), based on our present limited statistics.}
\label{fig:light}
\end{figure}

In Fig.~\ref{fig:light} we show the preliminary results for the $u$ and $d$-quark contributions to $a_\mu^{had}$ and $\delta a_\mu^{had} / a_\mu^{had}$, based on our present limited statistics.
The strong IB effect, due to the quark mass difference $(m_d - m_u)$ determined in Ref.~\cite{Giusti:2017dmp}, has been included in $\delta a_\mu^{(u,d)} / a_\mu^{(u,d)}$.
It can be seen that our results for the ratio $\delta a_\mu^{(u,d)} / a_\mu^{(u,d)}$ are in the range $0 - 1 \%$ (see also Ref.~\cite{Boyle:2017gzv}).
The improvement of the statistics is ongoing.

\section*{Acknowledgements}
We warmly thank R.~Frezzotti, K.~Jansen, M.~Petschlies, G.C.~Rossi, N. Tantalo and C.~Tarantino for many fruitful discussions and comments. 
We gratefully acknowledge the CPU time provided by PRACE under the project Pra10-2693 and by CINECA under the initiative INFN-LQCD123 on the BG/Q system Fermi at CINECA (Italy).

\end{document}